\documentclass[12pt]{article}
\usepackage{epsfig,url}

\def\beq{\begin{equation}}
\def\eeq#1{\label{#1}\end{equation}}
\def\eeqn{\end{equation}}
\def\beqa{\begin{eqnarray}}
\def\eeqa#1{\label{#1}\end{eqnarray}}
\def\eeqan{\end{eqnarray}}
\def\CR{\nonumber \\ }
\def\leqn#1{(\ref{#1})}

\def\psip{\psi^\prime}
\newcommand\iden{\leavevmode\hbox{\small1\normalsize\kern-.33em1}}

\def\stacksymbols #1#2#3#4{\def\theguybelow{#2}
    \def\vp{\lower#3pt}
    \def\sp{\baselineskip0pt\lineskip#4pt}
    \mathrel{\mathpalette\intermediary#1}}

\def\intermediary#1#2{\vp\vbox{\sp
     \everycr={}\tabskip0pt
     \halign{$\mathsurround0pt#1\hfil##\hfil$\crcr#2\crcr
              \theguybelow\crcr}}}

\def\lapproxeq{\stacksymbols{<}{\sim}{2.5}{.2}}

\begin{document}

\title{Little Higgs Models and T Parity}
\author{Maxim Perelstein \\
{\it Cornell Institute of High-Energy Phenomenology}\\
{\it Cornell University, Ithaca, NY 14853, USA}\\
{\tt E-mail: maxim@lepp.cornell.edu}} 
\date{May 12, 2006}
\maketitle

\begin{abstract}
Little Higgs models are an interesting extension of the Standard Model at the
TeV scale. They provide a simple and attractive mechanism of electroweak 
symmetry breaking. We review one of the simplest models of this class, the
Littlest Higgs model, and its extension with T parity. The model with T parity 
satisfies precision electroweak constraints without fine-tuning, contains an
attractive dark matter candidate, and leads to interesting phenomenology at 
the Large Hadron Collider (LHC).  
\end{abstract}

\section{Introduction}

In this contribution, we will briefly review the Little Higgs (LH) models, an interesting new class of theories of electroweak symmetry breaking (EWSB) that recently attracted considerable attention. For a more comprehensive introduction, the reader is referred to two recent review articles on this subject~\cite{STSrev,Prev}.

Precision electroweak data prefer a light Higgs boson: $m_h\lapproxeq 245$ GeV at 95\% c.l., assuming no other new physics~\cite{PDG}. A satisfactory theory of EWSB must contain a mechanism to stabilize the Higgs mass against radiative corrections. One intriguing possibility is that the Higgs is a composite particle, a bound state of more fundamental constituents held together by a new strong force~\cite{composite}. This scenario relates the weak scale to the confinement scale of the new strong interactions, which is generated via dimensional transmutation and can be naturally hierarchically smaller than the Planck scale. However, since precision electroweak data rule out new strong interactions at scales below about 10 TeV, an additional mechanism is required to stabilize the ``little hierarchy'' between the Higgs mass and the strong interaction scale. In analogy with the pions of QCD, the lightness of the Higgs could be explained if it were a {\it
Nambu-Goldstone boson} (NGB) corresponding to a spontaneously broken global symmetry of the new strongly interacting sector. Gauge and Yukawa couplings of the Higgs, as well as its self-coupling, must violate the global symmetry explicitly: an exact NGB only has derivative interactions. Quantum effects involving these interactions generate a mass term for the Higgs. In a generic model, the dominant effect comes from one-loop quadratically divergent part of the Coleman-Weinberg (CW) potential, and its large size makes the models phenomenologically unacceptable: either the Higgs is too heavy to fit the data, or the strong coupling scale is too low. Little Higgs models avoid this difficulty by incorporating the ``collective symmetry breaking''\footnote{The collective symmetry breaking mechanism was first gleaned by Arkani-Hamed, Cohen and Georgi~\cite{bigmoose} from a study of five-dimensional theories through the application of the dimensional deconstruction approach~\cite{Decon}.} mechanism, which enforces the cancellation of the quadratically divergent one-loop contributions to the Higgs mass, making a light composite Higgs compatible with the 10 TeV strong interaction scale. 
The cancellation is due to a set of new TeV-scale particles (typically gauge bosons and vector-like quarks) predicted by the LH models. If these models are realized in nature, the LHC experiments should be able to discover these particles and study their properties extensively.

\section{Littlest Higgs Model}
 
Many LH models have been proposed in the literature; as an example, let us briefly review the "Littlest Higgs" model~\cite{LH}, which provides one of the most economical implementations of the idea and forms the basis for most phenomenological analyses. Consider a model with an $SU(5)$ global symmetry, spontaneously broken down to an $SO(5)$ subgroup, at a scale $f\sim 1$ TeV, by a vacuum condensate in the symmetric tensor representation:
\beq
\Sigma_0 = \left(\begin{array}{ccc}
0 & 0 & \iden\\                                    
0 & 1 & 0\\                                
\iden & 0 & 0 \end{array} 
\right),
\eeq{sigma0}
where $\iden$ is a 2$\times$2 identity matrix. 
The model contains 14 massless NGB fields $\pi^a$, one for each broken generator $X^a$. At energy scales below $\Lambda\sim 4\pi f$, the NGB interactions are independent of the details of the physics giving rise to the condensate and can be described by an $SU(5)/SO(5)$ non-linear sigma model (nl$\sigma$m), in terms of the sigma field $\Sigma(x)=e^{2i\Pi/f}\Sigma_0$, where $\Pi=\sum_a \pi^a(x) X^a$. An $[SU(2)\times U(1)]^2$ subgroup of the $SU(5)$ is weakly gauged. The gauged generators are embedded in such a way that gauging each $SU(2)\times U(1)$ factor leaves an $SU(3)$ subgroup of the global symmetry unbroken:
\beqa
&Q_1^a=\left( \begin{array}{ccc} \sigma^a/2 &0 & 0 \\
0 & 0 & 0\\ 0 & 0 & 0
\end{array}\right), \ \ \ &Q_2^a=\left( \begin{array}{ccc} 0 & 0 & 0\\
0 & 0 & 0 \\
0 &0&-\sigma^{a*}/2\end{array} \right),\nonumber \CR 
&Y_1=
{\rm diag}(3,3,-2,-2,-2)/10\,, & Y_2={\rm
diag}(2,2,2,-3,-3)/10~.
\eeqa{gauged}
At the scale $f$, the condensate $\Sigma_0$ breaks the full gauge group down to the diagonal $SU(2)\times U(1)$, identified with the SM electroweak group. Four gauge bosons, $W_H^\pm, W_H^3$ and $B_H$, acquire TeV-scale masses by absorbing four of the NGB fields. The remaining NGBs decompose into a weak doublet, identified with the SM Higgs $H$, and a weak triplet $\Phi$:
\beq
\Pi = \left(\begin{array}{ccc}
* & H & \Phi\\                                    
H^\dagger & * & H^T\\                                
\Phi & H^* & * \end{array} 
\right),
\eeq{NGBs}
where asterisks denote eaten fields.
At the quantum level, gauge interactions induce a Coleman-Weinberg potential for the NGBs. However, the Higgs is embedded in such a way that the subset of global symmetries preserved by each $SU(2)\times U(1)$ gauge factor would be sufficient to ensure the exact vanishing of its potential. Both gauge factors, acting collectively, are needed to break enough symmetry to induce a non-zero CW potential for $H$: any diagram contributing to this potential must involve at least one power of a gauge coupling from each factor. One loop diagrams satisfying this criterion are at most logarithmically divergent; the usual one-loop quadratic divergence in the Higgs mass does not appear. The same collective symmetry breaking approach can be used to eliminate the large contribution to the Higgs mass from the top quark loops: the top Yukawa arises from two terms in the Lagrangian, each of which by itself preserves enough global symmetry to keep the Higgs exactly massless. Implementing this idea requires the introduction of a new vector-like fermion, the $T$ quark, with mass $m_T\sim f$ and the quantum numbers of the SM $t_R$. It is interesting that, in contrast to SUSY, the cancellations in the LH model involve particles of {\it the same spin}: the divergence due to the SM top loop is cancelled by $T$ loops, while the divergence due to the SM gauge bosons is cancelled by the loops of $W_H$ and $B_H$. The leading contribution to the CW potential from top loops has the form
\beq
   m_h^2 =  - 3 { \lambda_t^2 m_{T}^{2} \over 8\pi^2 }  
                   \log {\Lambda^2\over m_T^2} \ ,
\eeq{toploop}
and has the correct sign to trigger EWSB. 
The contributions from gauge and scalar loops have the opposite sign, but are typically smaller than~\leqn{toploop} due to the large top Yukawa; the two-loop contributions are subdominant. The triplet $\Phi$ is not protected by the collective symmetry breaking mechanism, and acquires a TeV-scale mass at one loop. An order-one 
Higgs quartic coupling is also generated, both by quadratically divergent one-loop diagrams and by tree-level exchanges of the triplet $\Phi$ arising from the 
$H^\dagger\Phi H$ vertex present in the one-loop CW potential. Thus, the model provides an attractive picture of radiative EWSB, with the required hierarchies $v\sim f/(4\pi) \sim \Lambda/(4\pi)^2$ emerging naturally.

The Littlest Higgs model is remarkably predictive, describing the TeV-scale new physics with only a small number of free parameters. The model contains two $SU(2)$ gauge couplings, two $U(1)$ couplings, and two couplings in the top Yukawa sector; however, in each case, one combination of the two is fixed by the requirement to reproduce the SM $g$, $g^\prime$, and $y_t$. This leaves three independent parameters; it is convenient to use three mixing angles, $\psi$, $\psip$, and $\alpha$, respectively. These angles, along with the scale $f$, determine the masses and couplings of the new states; for example,
\beq
M(W_H) \,=\, \frac{g}{\sin 2\psi}\,f\,,~~
M(B_H) \,=\, \frac{g^\prime}{\sqrt{5} \sin 2\psip}\,f\,,~~M(T)\,=\, \frac{\sqrt{2}\lambda_t}{\sin 2\alpha}\,f.
\eeq{masses}
Two additional parameters, coefficients $a$ and $a^\prime$ from the quadratically divergent part of the one-loop CW potential, are required to describe the weak triplet sector.

\begin{figure}[t]
\begin{center}
\includegraphics[width=10cm]{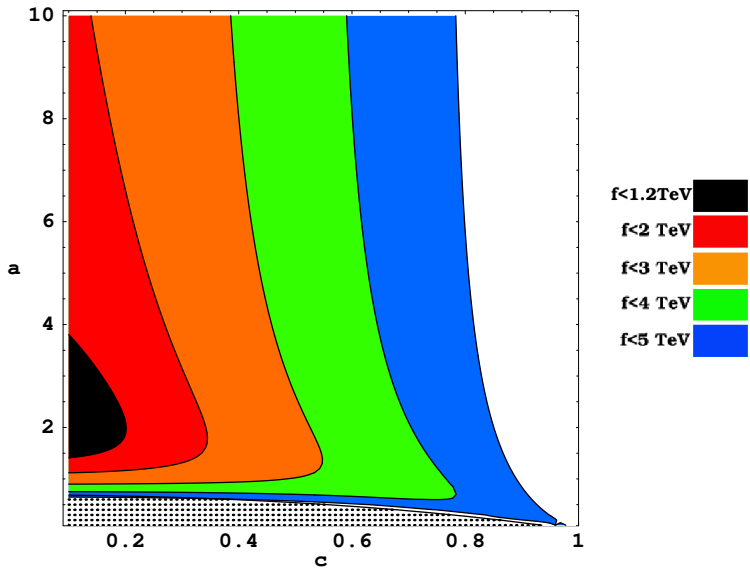}
\vskip2mm
\caption{Contour plot of the allowed values of $f$ in the $SU(5)/SO(5)$ Littlest Higgs model with an $SU(2)\times SU(2)\times U(1)$ gauged subgroup, as a function of the parameters $c\equiv \cos \theta$ and $a$. The gray shaded region at the bottom is excluded by requiring a positive triplet mass. From Ref.~\cite{CC2}.}
\label{fig:oneU1}
\end{center}
\end{figure}

The new particles at the TeV scale introduced by the Littlest Higgs model affect precision electroweak observables, and their properties are constrained by data. These constraints have been worked out in detail in Refs.~\cite{CC1,PEW}. Unfortunately, it was found that the simplest version of the model outlined above is strongly disfavored by data: the symmetry breaking scale is bounded by $f > 4$ TeV at 95\% c.l. in the ``best-case" scenario, and the bound is even stronger for generic parameters. Such a high $f$ would require a substantial amount of fine tuning to maintain the lightness of the Higgs, largely destroying the original motivation for the Littlest Higgs model. The corrections to observables are predominantly generated by the tree-level exchanges of heavy gauge bosons and the non-zero vacuum expectation value (vev) of the weak triplet $\Phi$; both these effects violate the custodial $SU(2)$ symmetry. The gauge boson contribution is dominated by the $B_H$, whose mass is typically well below the scale $f$, see Eq.~\leqn{masses}. The simplest way to alleviate the situation is to reduce the gauge group to $SU(2)\times SU(2) \times U(1)_Y$, abandoning the collective symmetry breaking mechanism in the $U(1)$ sector. This eliminates the $B_H$ boson, and consistent fits for $f$ as low as 1 TeV can be obtained~\cite{CC2,PPP}, albeit only in a rather small region of the parameter space as shown in Fig.~\ref{fig:oneU1}. Due to the small value of the SM $U(1)_Y$ coupling, the uncanceled contribution to the Higgs mass from this sector does not introduce significant fine tuning. While technically natural, this model seems rather unattractive from the theoretical point of view. 

\section{Littlest Higgs with T Parity}

A more elegant solution to the precision electroweak woes of the Littlest Higgs model has been proposed by Cheng and Low~\cite{LHT0}. They enlarge the symmetry structure of the models by introducing an additional discrete symmetry, dubbed ``T parity'' in analogy to R parity in the minimal supersymmetric standard model (MSSM). T parity can be implemented in any LH model based on a product gauge group, including the Littlest Higgs~\cite{LHT1,JP}. The gauge and higgs sector of the Littlest Higgs model with T parity (LHT) are identical to the Littlest Higgs model described above, with the additional restrictions $\psi=\psi^\prime=\pi/4$. (The action of the T parity on the gauge sector simply interchanges the two $SU(2)\times U(1)$ factors, enforcing the equality of the gauge couplings.) The SM gauge bosons are T-even, while their heavy counterparts $W_H$ and $B_H$ are T-odd. Likewise, the SM Higgs boson is T-even, while the additional weak-triplet scalar is T-odd. In the fermion sector, the parity is implemented in such a way that all the SM fermions are T-even. Consistency requires an introduction of a vector-like T-odd partner for each weak doublet fermion of the SM: T-odd quarks $\tilde{Q}^a_i$ are the partners of the quark doublets $Q^a_i$ (here $a=1\ldots 3$ is the color index, while $i=1\ldots 3$ is the generation index), while the T-odd leptons $\tilde{L}_i$ are the partners of the SM lepton doublets $L_i$. In the top sector, the $T$ quark of the Littlest Higgs model is T even, and an additional T-odd top partner $T_-$ has to be introduced for consistency.

\begin{figure}[tb]
\centerline{\includegraphics[width=0.5\hsize]{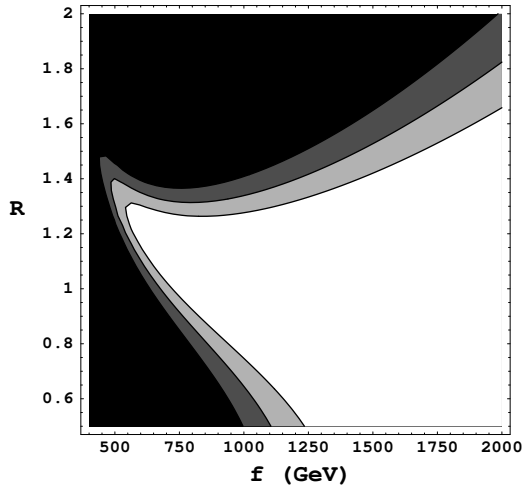}}
\vskip-0.3cm
\caption{Exclusion contours in terms of the parameter $R\equiv\tan\alpha$ 
and the symmetry breaking scale $f$ in the LHT model. The contribution of the T-odd fermions 
to the T parameter is neglected. From lightest to darkest, the contours 
correspond to the 95, 99, and 99.9 confidence level exclusion.
From Ref.~\cite{HMNP}.} 
\label{fig:TparEW}
\end{figure}

From these assignments, it is easy to see that the T parity explicitly forbids any tree-level contribution from the heavy gauge bosons to the observables involving only SM particles as external states, as well as the $H^\dagger\Phi H$ coupling in the effective CW potential which  induced the triplet vev in the original Littlest Higgs model. Thus the leading, tree-level corrections to precision electroweak observables in the Littlest Higgs model are eliminated. In fact, T parity guarantees that corrections to all observables involving only the SM external states are generated exclusively at loop level.\footnote{The only exception occurs in the top sector, where the SM top quark and the $T$ mix at the tree level resulting to deviations of the top couplings from their SM values. At present, this does not constrain the model, due to the limited experimental precision on observables involving the top quark. However, these effects are likely to become observable at the LHC and the ILC~\cite{Snowmass}.}
A detailed analysis of precision electroweak constraints on the LHT model at the one-loop level has been performed in Ref.~\cite{HMNP}. It was found that values of $f$ as low as 500 GeV are allowed, as illustrated in Fig.~\ref{fig:TparEW}. In addition, experimental constraints on the coefficients of four-fermion operators such as $\bar{e}e\bar{u}u$ place an {\it upper} bound on the mass of the T-odd quark and lepton partners. Assuming a common mass scale $\tilde{M}$ for all partners, Ref.~\cite{HMNP} obtained 
\beq
\frac{\tilde{M}}{1~{\rm TeV}} < 4.8 \left(\frac{f}{1~{\rm TeV}}\right)^2\,.
\eeq{upperT}

\begin{figure}[tb]
\begin{center}
\includegraphics[width=6.5cm]{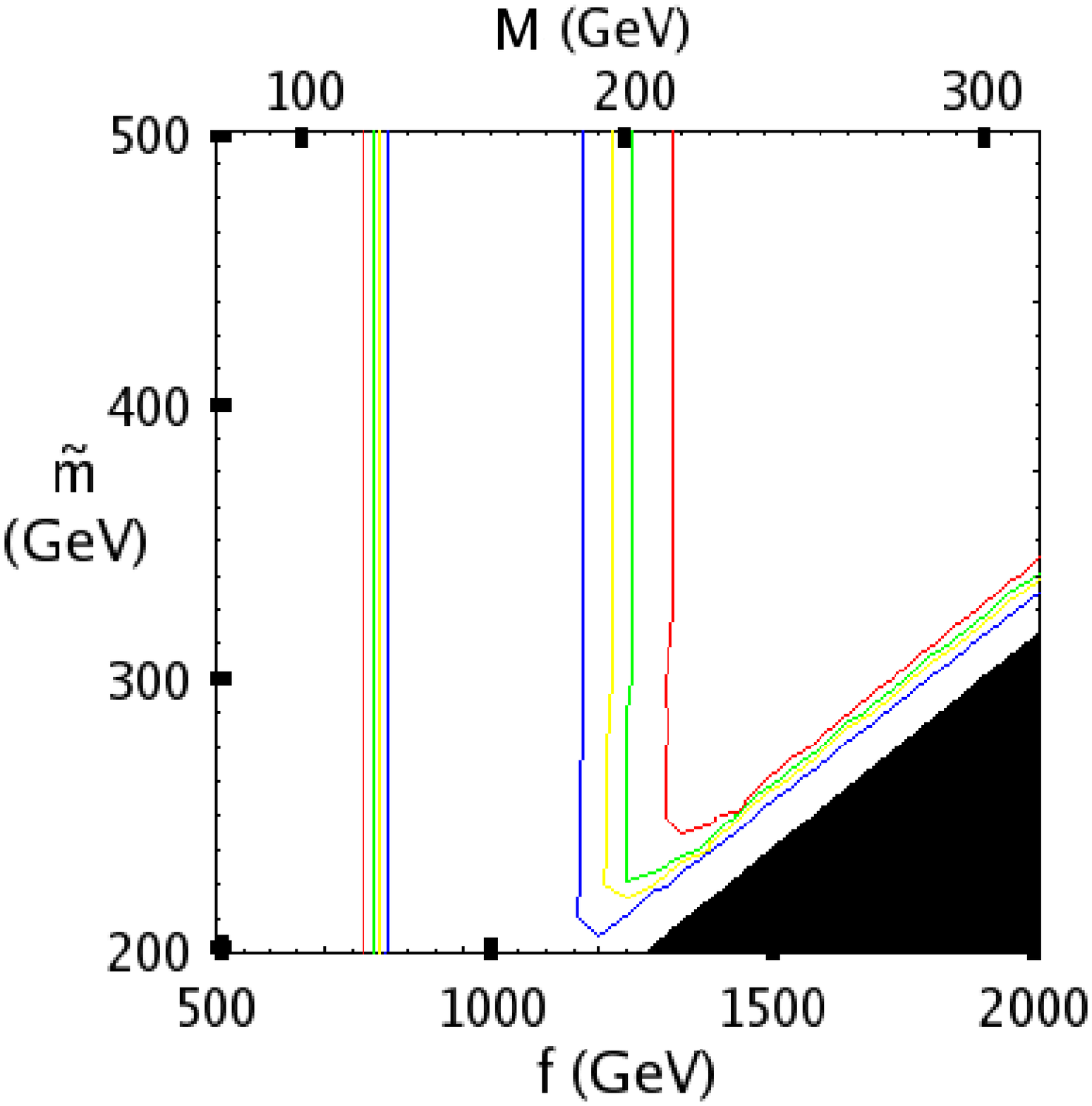}
\includegraphics[width=6.5cm]{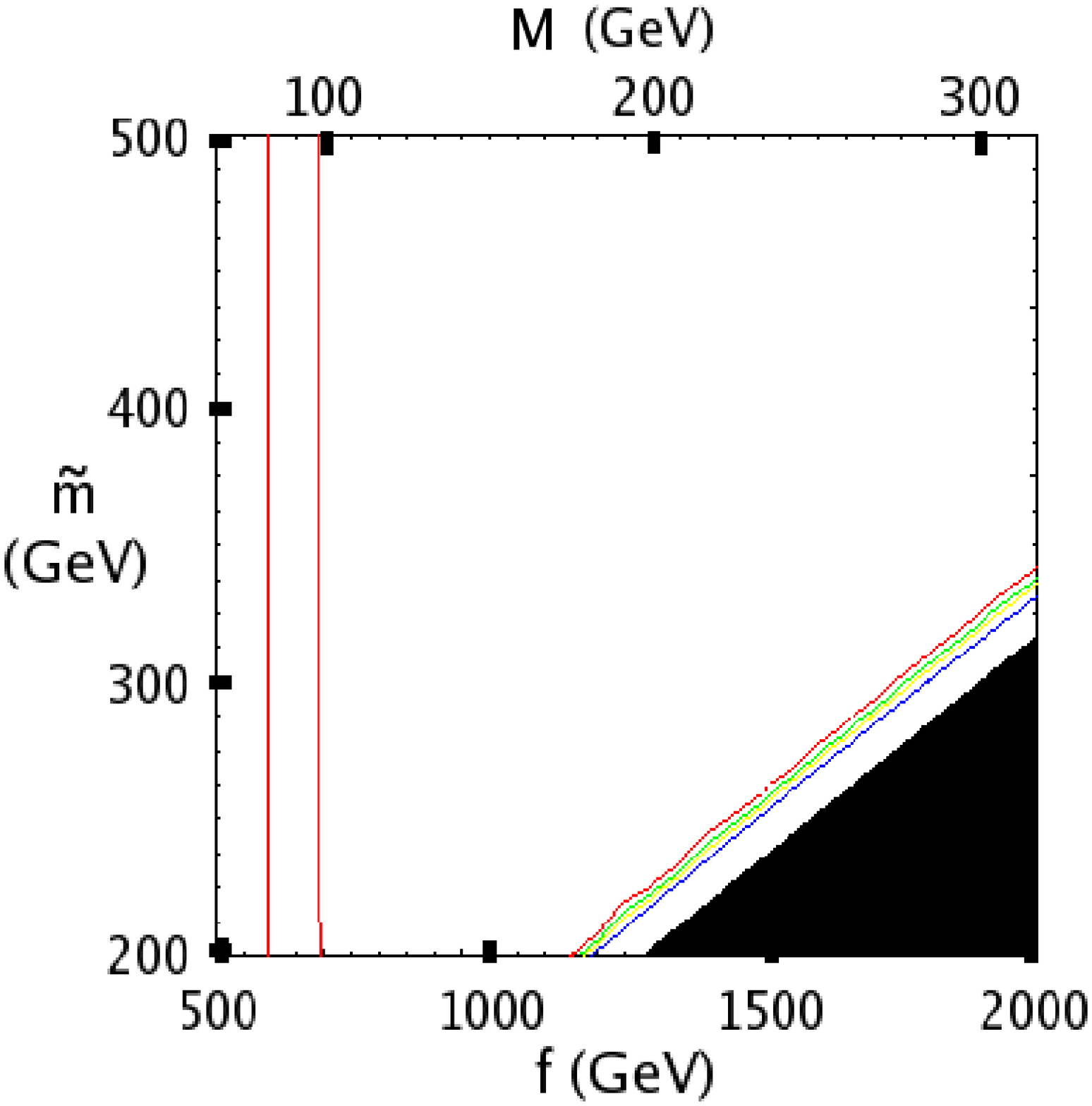}
\vskip2mm
\caption{The contours of constant present abundance of the heavy photon
LTP, $\Omega_{\rm LTP} h^2$, in the $M-\tilde{M}$ plane, where $M$ is the LTP mass and $\tilde{M}$ is the common mass scale of the T-odd quark and lepton partners. The Higgs mass is 
taken to be 300 GeV (left panel) and 120 GeV (right panel). The red and green 
contours correspond to the upper and lower bounds from the first-year WMAP data, 
assuming that the LTP makes up all of dark matter. The yellow 
and blue lines correspond to the LTP contributing 50\% and 70\%,
respectively, of the measured dark matter density. The shaded 
region corresponds to a charged and/or colored LTP. From Ref.~\cite{BNPS}.}
\label{fig:DM}
\end{center}
\end{figure}

In analogy to the lightest supersymmetric particle (LSP) of SUSY models with R parity, the lightest T-odd particle (LTP) of the LHT model is stable. Typically, the 
LTP is the $B_H$ gauge boson, frequently referred to as the "heavy photon". Indeed, this particle is quite light, $M(B_H)=g^\prime f/\sqrt{5}\approx 0.16 f$. Since the $B_H$ is weakly interacting, it provides a potential dark matter candidate. 
The relic abundance of the heavy photon has been computed in Refs.~\cite{JP,KEK}; the most complete calculation to date, which included coannihilation effects, was recently presented in Ref.~\cite{BNPS}. Regions of parameter space where the relic photon has the correct density to account for the observed dark matter have been mapped out; see, for example, Fig.~\ref{fig:DM}. As expected, the correct relic density can be obtained for reasonable values of the model parameters, consistent with naturalness and other constraints.
Prospects for direct detection of the heavy photon dark matter have been evaluated in Ref.~\cite{BNPS}; unfortunately the rates are quite low and a factor of $10^3$ improvement of the current sensitivity of the CDMS experiment is required before the interesting regions of the parameter space can be probed. Indirect detection via anomalous high-energy positrons~\cite{KEK} and gamma rays~\cite{BNPS} arising from the pair-annihilation of galactic LTPs provides a more promising detection avenue, with observable signals expected at the PAMELA, AMS-02, and GLAST experiments. 

A first study of the collider phenomenology of the Littlest Higgs model with T parity was presented in Ref.~\cite{JP}. Most signatures are characterized by large missing energy or transverse momentum carried away by the two LTPs. In this sense, the signatures are very similar to SUSY models with conserved R parity or UED models with  conserved Kaluza-Klein parity, raising an interesting question of how these models can be distinguished experimentally at the LHC and the ILC. This question deserves future study. In addition, the T-even heavy top $T_+$ can be produced singly. The phenomenology of this particle has been studied extensively in the context of the original Littlest Higgs~\cite{Han,PPP}. A detector-level study by the ATLAS collaboration~\cite{ATLAS} concluded that the discovery reach for the $T_+$ at the LHC is about 2 TeV. In the model with T parity, the $T_+$ may acquire additional decay channels~\cite{JP}; however, for generic parameters, the analyses performed in~\cite{Han,PPP,ATLAS} are still applicable.

\section{Conclusions}

In this contribution, we have briefly reviewed the Little Higgs models, which provide an attractive scenario combining dynamical stabilization of the weak-Planck hierarchy by dimensional transmutation with the radiative EWSB. We concentrated on the Littlest Higgs model; for a review of many other LH models in the literature, the interested reader is referred to the review articles~\cite{STSrev,Prev}. The version of the model incorporating T parity provides acceptable fits to precision electroweak data without significant fine tuning. The model contains an interesting and phenomenologically successful dark matter candidate, the lightest T-odd particle (LTP). It makes interesting predictions which can be tested at the LHC. More work is required in order to ensure that the LHC experiments maximize their potential in searching for the predicted signatures; to this end, it would be useful to systematically incorporate the LH model into the standard Monte Carlo packages such as {\tt MadGRAPH}.

{\bf Acknowledgment ---} This research is supported by the
National Science Foundation grant PHY-0355005.

\end{document}